# Viscosity, entropy and the viscosity to entropy density ratio; how perfect is a nucleonic fluid?

Aram Z. Mekjian

Department of Physics and Astronomy, Rutgers University, Piscataway, NJ 08502,
TRIUMF Laboratory, 4004 Westbrook Ave. , Vancouver, B.C. CA  V6T 2A3

Abstract

The viscosity of hadronic matter is studied using a classical evaluation of the scattering angle and a quantum mechanical discussion based on phase shifts from a potential. Semi classical limits of the quantum theory are presented. A hard sphere and an attractive square well potential step are each considered as well as the combined effects of both. The lowest classical value of the viscosity for an attractive potential is shown to be a hard sphere limit. The high wave number-short wavelength limits of the quantum result have scaling laws associated with it for both the viscosity and entropy. These scaling laws are similar to the Fraunhofer diffraction increase for the hard sphere geometric cross section. Specific examples for nuclear collisions are given. The importance of the nuclear tensor force and hard core is mentioned. The viscosity $\eta$, entropy density $s$ and $\eta/s$ ratio are calculated for a gas of dilute neutrons in the unitary limit of large scattering length. Away from the unitary limit, the ratio of the interaction radius or the scattering length to the interparticle spacing introduces a variable $y$ besides the fugacity $z$. The isothermal compressibility is shown to impose important constraints. The results for $\eta/s$ are compared to the AdS/CFT string theory minimum of $(1/4\pi)\hbar/k_B$ to see how close a nucleonic gas is to being a perfect fluid. The $\eta/s \sim 1\,\hbar/k_B$ for a neutron gas in its unitary limit. The $\eta/s \sim 3\,\hbar/k_B$ treating the nuclear scattering as billiard ball collisions. The minimum $\eta/s$ for a neutron gas occurs in regions of negative isothermal compressibility and high fugacity where higher virial terms are important. In a neutron-proton system higher virial terms are associated with a liquid-gas phase transition and critical opalescent phenomena. A connection between the nuclear flow tensor and viscosity is developed using a Fokker-Planck equation and a relaxation time description. The type of flow-laminar, vortex, turbulent- is investigated.

## I. Introduction

Understanding the viscosity of matter is of interest in many regimes of energy and in several different areas of physics. These areas include atomic systems, nuclear matter, neutron star physics, low energy to relativistic energy heavy ion collisions and at the extreme end string theory. For example viscosity plays a role in collective flow phenomena in medium and relativistic heavy ion collisions (RHIC). Specifically, viscosity inhibits or resists flow in such collisions. Ref. [1,2] are early theoretical studies of viscosity and flow. Some recent experimental results for RHIC physics can be found in Ref. [3] and some overviews are in Ref. [4,5]. A microscopic kinetic theory description of viscosity involves the transport of momentum across an area and involves particle interactions through mean free path arguments. If interactions are strong, the shear viscosity is small. The shear viscosity is inversely proportional to the scattering cross section. Substances with low kinematic viscosity and with high Reynolds number do not flow smoothly as in laminar flow, but rather form eddies when flowing around obstacles. The Reynolds number also depends on the flow velocity and the eddy behavior is easily seen by putting a paddle in water in a boat moving at various speeds. Surprisingly, in relativistic heavy ion collisions, the quarks and gluons act as a strongly coupled liquid [6,7,8] with low viscosity rather than a nearly ideal gas of asymptotically free particles with high viscosity. Low viscosity fluids which interact strongly are called nearly perfect fluids. String theory has put a small lower limit on the ratio of shear viscosity $\eta$ over entropy density $s$ given by $\eta/s \geq \hbar/(4\pi k_B)$ [9] with $k_B$ the Boltzmann constant. The string theory result has generated considerable interest in questions concerning strongly correlated systems and viscosity. The focus of the present paper is on low and moderate energy systems of interacting nucleons. A system of nucleons has parallel properties such as correlated behavior, bound states, phase transitions and critical point behavior.

Nearly perfect fluids with very low viscosity appear also in cold atoms. Properties of interacting quantum degenerate Fermi gases were first observed in atomic systems [10-12]. Atoms are cooled in a laser trap to the point where quantum statistics and an associated thermal wavelength play an important role. Such systems can be tuned by using a magnetic field and properties of Feshbach resonances are used to study the strong coupling crossover from a Bose–Einstein condensate of bound pairs to a Bardeen-Cooper-Scrieffer BCS superfluid state of Cooper pairs. A remarkable aspect of strongly interacting Fermi gases is a universal behavior which occurs when the scattering length is very large compared to the interparticle spacing. In this unitary limit, properties of a heated gas are determined by the density $\rho$ and temperature $T$, independent of the details of the two body interaction. Early theoretical discussions of dilute Fermi systems at infinite scattering length can be found in Ref. [13,14]. At temperature $T = 0$, the Fermi energy $E$ of a strongly interacting Fermi gas differs from the Fermi energy $E_F$ of a non-interacting Fermi gas by a universal factor $\xi$ with $E = \xi E_F$. Accounting for this difference in nuclear systems is referred to as the Bertsch challenge problem [15]. Initial work on this problem was done by Barker [16] and latter Heiselberg [17]. A Monte Carlo numerical study of the unitary limit of pure neutron matter is given in Ref. [18]. Analytic studies of pure neutron systems can be found in the extensive work of Bulgac and collaborators [19-21]. In these studies the dimensionless factor $\xi \approx 0.4$.

Part of present paper is an extension of an earlier work [1] in which the viscosity of hadronic matter was studied using a relaxation time approximation to the Boltzmann equation and also a Fokker-Planck description. This earlier paper focused on features associated with collective flow. Specifically, the kinetic flow tensor was related to the pressure tensor and the collective velocity field. The pressure tensor, in turn, was related to the nuclear viscosity. The Kubo-Green formulae [22,23,24] relates the viscosity to time fluctuations in the pressure tensor. A calculation of the Reynolds number showed that the flow is laminar. A Fokker Planck approach has been recently used [25] to study the viscosity of the quark-gluon phase. A relaxation time approach also appears in Ref. [26,27] for trapped Fermi gas in a oscillator well near the unitary limit of large scattering length. Questions related to entropy in nuclear systems were also studied early on [28] using the Sakur-Tetrode law and the Saha-equation. Recently, the behavior of entropy in the unitary limit was given [29]. The work presented in this paper gives a more refined calculation of the viscosity than given in Ref. [1] as well as further discussions of the entropy. Both classical and a quantum approaches to the viscosity are discussed using a Chapman-Enskog description [30,31]. The classical calculation involves the scattering angle while the quantum approach relates the viscosity to properties of the scattering phase shifts. In such studies a potential must be specified. The interactions used here to describe the nuclear scattering are a square well potential, a pure hard core and a combination of the two. The pure hard core potential pictures the collisions as arising from impenetrable billard balls and is used as a comparison. The discussion of viscosity from a pure hard core potential also contains new results which show a scaling law regime with increasing momentum of the colliding pair of nucleons. This scaling feature parallels results that arise when considering scattering from a hard sphere [32] which lead to a diffractive increase by a factor of two from the classical result of $\pi R^2$. The square well potential with a hard core approximates a nuclear potential and the role of the nuclear hard core as well as the nuclear tensor force are mentioned. A study of viscosity using a delta-shell potential can be found in Ref.[33,34]. The focus here will be on a pure one component system- such as in a gas of neutrons. One important feature of the interaction between nucleons is the very large scattering length $a_{sl}$ and its associated unitary regime. The unitary limit of the viscosity and entropy is examined for this system. A study of Feshbach resonances and the second virial coefficient in atomic systems is given in ref. [35,36]. Viscosity to entropy considerations also appear in the damping of giant resonances in nuclear physics [37] and in atoms in laser traps [38].

## II. Classical and quantum approaches to the viscosity

### II. A. Mean free path and relaxation time approaches

First, the standard discussion of the viscosity that can be found in textbooks such as Ref. [39] relate the viscosity to the number density $\rho$, mass of a fluid particle $m$, mean speed $\hat{v} = \sqrt{8k_B T / m\pi}$ of a Boltzmann distribution and mean free path $l_\lambda$ as

$$\eta = \frac{1}{3} \rho m \hat{v} l_\lambda . \qquad (1)$$

The mean free path $l_\lambda = 1/(\rho\sigma)$. The scattering cross section $\sigma = \pi D^2$ for hard sphere scattering of particles with diameter $D$. For this description the viscosity $\eta$ no longer depends on the number density $\rho$ and is simply

$$\eta = \frac{1}{3} m\hat{v} \frac{1}{\pi D^2} . \qquad (2)$$

The $\eta$ does not involve Planck's constant. If the mean free path is taken as $l_\lambda = \hat{v}\tau$, with $\tau$ the mean time between collisions, the $\eta \sim \varepsilon\tau$. The $\varepsilon = E/V \sim m\hat{v}^2\rho$ is the energy density and $E$ is the energy per particle. The $\rho$ now explicitly appears in $\eta$ but is removed as follows. Taking the entropy density as $s \sim \rho k_B$, the ratio $\eta/s \sim E\tau/k_B$ is again $\rho$ independent. The product $E\tau$ is governed by an uncertainty relation [2,4,5] in a quantum approach, giving $\eta/s \sim \hbar/k_B$ with the particular constant of proportionality $1/4\pi$ from string theory. In a relaxation time approximation to the Boltzmann equation [1], the collision term is replaced with $(df/dt)|_{coll} = -(f-f^0)/\tau_R$ with $f^0$ the local equilibrium phase space distribution. The viscosity is then $\eta = \rho\tau_R k_B T$ and is directly related to a relaxation time $\tau_R$. The $m\hat{v}^2 \sim k_B T$ which leads to $\eta \sim \rho E \tau_R$ as discussed above.

### II.B. Chapman-Enskog Theory

The Chapman-Enskog theory[30,31] relates the viscosity $\eta$ to terms involving the scattering angle $\chi$ or phase shift $\delta_l$. Fig. 1 shows scattering off various potentials.

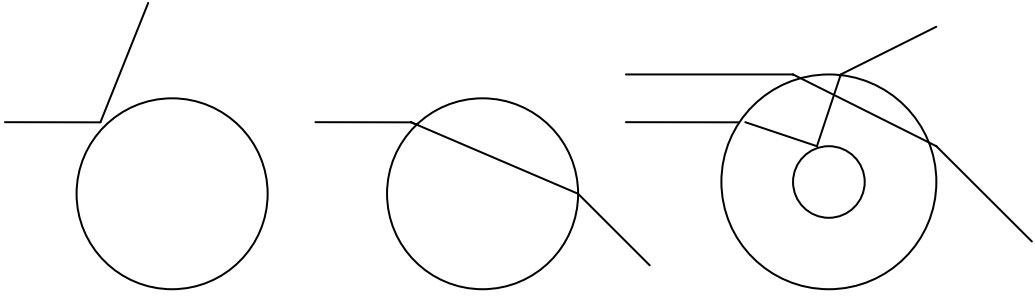

FIG.1. Scattering off various potentials. Left figure is hard sphere scattering. Middle figure is scattering off an attractive potential. Snell's law of reflection applies to the hard sphere case and gives the scattering angle $\chi$ or angle of deflection with respect to the initial direction as $\chi = \pi - 2\theta_i$. Snell's law of refraction $\sin\theta_i = n\sin\theta_f$ applies to the attractive well and has $\chi = 2(\theta_f - \theta_i)$. Right figure is the combined effect of an attractive square well with an inner hard core. The right figure is discussed in section II.B.5.

Angular momentum conservation leads to Snell law of refraction $n\sin\theta_f = \sin\theta_i$, with $\sin\theta_i = b/R$, and

$$n = n(E, V_0) = \sqrt{(E+|V_0|)/E} = \sqrt{1+\frac{2\mu|V_0|}{\hbar^2 k^2}} = \sqrt{1+\frac{2\mu|V_0|}{h^2}\lambda^2} \ . \tag{3}$$

The $b$ is the impact parameter and $E$ is the incident energy. This equivalence with an index of refraction shows why the reflection and refraction $n(E, V_0)$ of light can be explained in terms of both Huygens wave picture and classical Newtonian mechanics. Fermat's principle that the path taken is the one of least time leads to Snell's equations for reflection and refraction. The particle/quantum picture is due to Feynman and represents the process as a sum of all phasors over all possible paths that a photon will take in the case of light. The Fresnel equations determine the reflected and refracted intensities by boundary conditions on the E&M waves at an interface. Corresponding, the phase shifts for particles are obtained in a similar manner from the wave function. The wavelength dependence given in Eq. 3 arises from the deBroglie relation $\lambda = h/p$.

The viscosity is obtained from [30,31]

$$\eta = \frac{5}{8} \frac{\sqrt{\pi k_B T m}}{\pi R^2 \omega^{(2,2)}} \tag{4}$$

with $\omega^{(2,2)}$ given by

$$\omega^{(2,2)} \equiv \omega = \frac{1}{\pi R^2} \int d\gamma \cdot e^{-\gamma^2} \gamma^7 \phi_{12}^{(2)}(E, V, R) \tag{5}$$

where $\phi_{12}^{(2)}(E, V, R) \equiv \phi$ is evaluated both classically and in a quantum approach. The $\gamma^2 = E/k_B T$ with $E = \hbar^2 k^2 / 2\mu$ and with $\mu$ the reduced mass. The classical approach uses

$$\phi = 2\pi \int_0^\infty (\sin^2 \chi) b \cdot db. \tag{6}$$

The quantum calculation evaluates $\phi$ using

$$\phi = \frac{4\pi}{k^2} \sum_l \frac{(l+1)(l+2)}{2l+3} \sin^2(\delta_{l+2} - \delta_l). \tag{7}$$

The $\delta_l$ is the $l$'th phase shift of the potential used to describe the scattering. For identical particles, the factor $4\pi$ is replaced with $8\pi$ and the sum is over even $l$ for bosons and odd $l$ for "spinless" fermions. Spin effects for fermions can also be included – see appendix A. The total cross section for scattering is simply

$$\sigma = \frac{4\pi}{k^2} \sum_l (2l+1) \sin^2 \delta_l \tag{8}$$

and this expression has some features that parallel the expression for $\phi$. The cross section appears in Eq. 1 for the viscosity through the factor involving the mean free path. A somewhat related quantity that will appear in expressions for the entropy [29] is the Beth-

Ulhenbeck continuum integral [40-42] which is labeled $B_C$ and given by :

$$B_C = \frac{1}{\pi}\sum_l (2l+1)\int \frac{d\delta_l}{dk}\exp(-\hbar^2 k^2/mk_B T)dk . \qquad (9)$$

### II. B.1 Classical calculation of viscosity for a hard sphere potential.

For hard sphere of radius $R_C$ scattering, the impact parameter $b = R_C \cos\chi/2$ and $bdb = -(1/4)R_C^2(\sin\chi)d\chi$. Thus $\phi = 2\pi R_C^2/3$ and therefore $\omega^{(2,2)} = 2$. The viscosity is

$$\eta = \frac{5}{16}\frac{\sqrt{\pi k_B T m}}{\pi R_C^2}. \qquad (10)$$

This expression for the viscosity can be compared to that given by Eq. 1 with $\sigma = \pi R_C^2$ and $\hat{v} = \sqrt{8k_B T/m\pi}$ leading to

$$\eta = \frac{1}{3}\rho m\hat{v}l_\lambda = \frac{1}{3}\sqrt{\frac{8}{\pi^2}}\frac{\sqrt{\pi k_B T m}}{\pi R_C^2} . \qquad (11)$$

The ratio of these two expressions for $\eta$ is $(5/16)/(\sqrt{8}/3\pi)=1.04$, a difference of 4%. The differential cross section is $\sigma(\chi) = -(b/\sin\chi)\cdot db/d\chi = R_C^2/4$ and total $\sigma = \pi R_C^2$ which is the geometric cross section since anything hitting the sphere is scattered.

### II.B.2 Classical calculation of viscosity for a square well potential.

For a square well $\theta_f = \theta_i + \chi/2$, $\sin\phi_f = \sin\theta_i/n$ and thus

$$\sin\chi = 2\frac{b}{nR}\sqrt{1-\frac{b^2}{n^2 R^2}}(1-2\frac{b^2}{R^2}) - 2\frac{b}{R}\sqrt{1-\frac{b^2}{R^2}}(1-2\frac{b^2}{n^2 R^2}) . \qquad (12)$$

The connection between $b$ and $\chi$ can also be written as

$$\frac{b}{R} = \frac{\sin\frac{\chi}{2}}{\sqrt{1+\frac{1}{n^2}-\frac{2}{n}\cos\frac{\chi}{2}}} = \frac{n\sin\frac{\chi}{2}}{\sqrt{1+n^2-2n\cos\frac{\chi}{2}}} . \qquad (13)$$

When the index of refraction $n \to 1$, $\chi \to 0$ and when $n \to \infty$, $\theta_f \to 0$, and $\chi \to -2\theta_i$.

The differential scattering cross section can be obtained from $\sigma(\chi) = b/\sin\chi |db/d\chi|$. Letting $z = \cos(\chi/2)$, $\sigma(\chi) = n^2 R^2 (nz-1)(n-z)/(4z(1+n^2-2nz)^2)$. The $\sigma(\chi)$ is constrained by $nz - 1 \geq 0$. The largest scattering angle $\chi$ is when $b = R$, $\sin\theta_i = 1$ and $\sin\theta_f = 1/n$. The $z = \cos(\chi/2) = \cos(\theta_r - \pi/2) = \sin\theta_r = 1/n$. At this point $nz - 1$ vanishes and $\sigma(\chi) = 0$. The $\sigma = \pi R^2$, which is the same $\sigma$ as that of a hard sphere. The viscosity based on Eq. 1 would then have $l_\lambda = 1/\rho\pi R^2$. However, the Chapman-Enskog approach requires an evaluation of $\phi_{12}^{(2)}(E, V_0, R) \equiv \phi$ and then $\omega$ to obtain $\eta$. The $\phi$ is

$$\phi/2\pi R^2 = (16 - 30n - 40n^2 + 20n^3 + 40n^4 + 4n^5 + 20n^7 - 30n^9)/120n^4 +$$

$$(15(n^2-1)(n^8-1)/120n^4)\log[(n+1)/(n-1)]. \tag{14}$$

Two limits can be considered. One limit is for $n$ near unity and the other for large $n$. For $n$ near unity, $n = 1 + \varepsilon$ with $\varepsilon$ small. Then $\phi = 2\pi R^2 \left(-\varepsilon^2 \{2\log\varepsilon + 2\log 2 + 3\} - \varepsilon^3/3\right)$ and thus $\phi, \omega \to 0$ if $n \to 1$ and $\eta \to \infty$. However in this limit of infinite viscosity, the concept of momentum transport from collisions between layers of fluid fails since the particles move back and forth between the endpoints defined by the moving walls of the container. In subsection II.B.5 a hard core will be include inside the attractive potential and $\phi, \omega$ will not longer go to zero. For large $n$, $\phi = 2\pi R^2 (1/3 - 2/(35n) - 1/(3n^2)...)$. The 1/3 term in the parenthesis gives the hard sphere result. The $\phi_{12}^{(2)}(E, V_0, R)$ can be substituted into the integral for $\omega$, using $n = \sqrt{1 + h_T/\gamma^2}$ with $h_T \equiv V_0/k_B T$, which in turn determines $\eta$. The general behavior of $\omega$ with $h_T = V_0/k_B T$ is shown in Fig. 2.

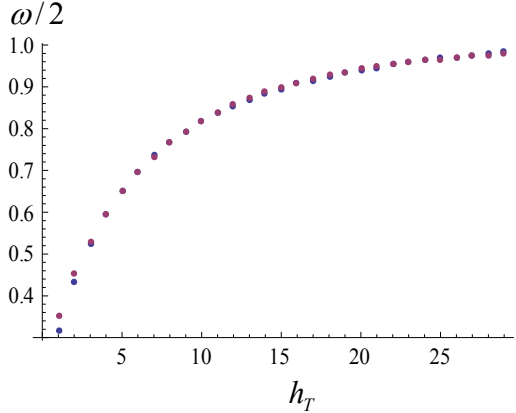

FIG. 2: (Color online) The classical behavior of $\omega/2$ versus $h_T = V_0/k_B T$ for an attractive potential. The limit $h_T \to \infty$ has $\omega/2 \to 1$ which is the hard sphere limit. The rise of $\omega/2$ to the value 1 is approximately exponential with $\omega \approx 2(1 - \exp(-0.25(h_T)^{0.8}))$. The exact calculation shown in the figure slightly overshoots 1 which will be neglected. The exponential representation has a slightly higher value at low $h_T$.

The results of Fig. 2 show that the classical calculation of the viscosity over a broad range of $h_T = V_0/k_B T$ from an attractive potential can be approximated as

$$\eta = \frac{5}{16} \frac{\sqrt{\pi k_B T m}}{\pi R^2 (1 - \exp(-0.25(V_0/k_B T)^{0.8}))} . \tag{15}$$

Therefore, in a classical evaluation of the viscosity, the smallest value of $\eta$ for an attractive interaction is at the largest value of $\omega$ which is the hard sphere result.

### II.B.3 Quantum calculation of viscosity for a hard sphere potential and the semi-classical limit $\hbar \to 0$.

First, results for hard sphere scattering will be given and compared to the classical evaluation. The phase shifts for a hard sphere are simply $\tan \delta_l = j_l(x)/\eta_l(x)$ with $j_l$ and $\eta_l$ Bessel functions. The $x = kR_C$, with $R_C$ the hard sphere radius and $k$ the wave number. The quantities $\phi, \omega, \eta$, as well as the entropy which will be used later, are all developed in appendix A. The $\omega$ can be rewritten as

$$\omega = 4\xi^4 \int_0^\infty dx \cdot e^{-\xi x^2} x^7 \left( \frac{1}{x^2} \sum_{l=0,1,2,\ldots} \frac{(l+1)(l+2)}{2l+3} \sin^2(\delta_{l+2}(x) - \delta_l(x)) \right) \tag{16}$$

with $\xi = (\lambda_T / R_C \sqrt{2\pi})^2$. The quantum wavelength $\lambda_T = h/\sqrt{2\pi k_B T m}$ is a de Broglie wavelength associated with the thermal momentum $E \sim k_B T \sim p^2/2m$. The $\gamma$ that appears in $\omega$ is $\gamma = (x/\sqrt{2\pi})(\lambda_T/R_C)$. The $\phi$ sum has the following scaling property when $x \to \infty$: $\phi \to 2\pi R_C^2/3$ which is the classical hard sphere scattering result mentioned above. The scaling behavior of $\phi(x)$ is shown in Fig.3. This scaling result parallels a similar result for the cross section which in the high energy limit $\sigma \to 2\pi R_C^2$. This factor of two increase over the hard sphere geometrical area $\pi R_C^2$ arises from diffraction.

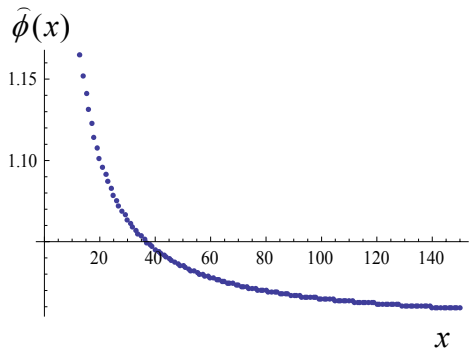

FIG. 3. (Color online) Scaling property of $\hat{\phi}(x)$ with $x$. The quantity plotted is $\hat{\phi} \equiv (6/x^2)\Sigma_l[(l+1)(l+2)/(2l+3)] \cdot \sin^2(\delta_{l+2}(x) - \delta_l(x))$ versus $x$. For this rescaled quantity the limiting value is unity. The $\phi = 4\pi R_C^2 \hat{\phi}/6$.

The other extreme energy is a low energy result. At very low energies only an $S$–wave phase shift is important. The cross section goes to $4\pi R_C^2$, or four times the geometrical result. Also, $\phi \to 8\pi R_C^2/3$ for $x \ll 1$. The $S$–wave phase shift is $\delta_0 = -kR_C = -x$, giving 2/3 for the bracket term in the integration for small $x$ since $\sin^2 x \approx x^2$. The integration of $\exp(-\xi x^2)x^7$ is $3/\xi^4$ resulting in $\omega = 8$ compared to the classical value $\omega = 2$. Corrections from spin and identical particles can be included and are given in appendix A. For now, these corrections will be neglected since a comparison is made with the classical calculation which does not contain these factors. The range of $\eta$ is then

$$\frac{5}{16}\frac{\sqrt{\pi k_B T m}}{\pi R_C^2} \geq \eta \geq \frac{5}{64}\frac{\sqrt{\pi k_B T m}}{\pi R_C^2} \; . \tag{17}$$

The behavior of $\omega$ is determined by $\xi = (\lambda_T/R_C\sqrt{2\pi})^2$ which in turn depends on the temperature $T$ through $\lambda_T$. Low $T$ has low associated energies and large $\lambda_T$. When $\lambda_T/R_C \gg 1$, only small $x$ contribute to the integral because $\exp(-\xi x^2)$ suppresses large $x$. Both extreme endpoints in Eq. (17) do not involve $\hbar$. The semi-classical limit has $\hbar \to 0$ and $\xi \to 0$. In this limit, the scaling behavior shown in Fig. 3 arises and $\eta$ is the classical value. To see how the viscosity evolves from the $S$–wave limit to the classical value, small $x$ expansions are made for the $P,D,F$ phase shifts as given in appendix A. These results for $\delta_0, \delta_1, \delta_2, \delta_3, ...$ can be used to obtain an expansion for $\omega$ for small $x$, further expanded in even $l$ and odd $l$ components $\omega_E$ and $\omega_O$, as

$$\omega = \left(8|_S - \frac{32}{3\xi}|_S + \frac{1}{\xi^3}(\frac{360}{7}|_{S,D} + \frac{256}{35}|_D)..\right)_E + \left(\frac{48}{\xi^2}|_P - \frac{1152}{5\xi^3}|_P + ..\right)_O \equiv \omega_E + \omega_O. \tag{18}$$

The factor 8 is the pure $S$–wave result. The $1/\xi^2$ term arises solely from a $P$–wave. The contribution of each partial wave is also given. It should be noted that the result of Eq. (18) gives an expansion for $\omega$ in inverse powers of $\hbar^2$ since $\xi = (\lambda_T/R_C\sqrt{2\pi})^2 \sim \hbar^2$. The viscosity is connected to this series expansion around the $S$–wave scattering limit using Eq. (4). The hard sphere quantum result for $\eta$ is shown in Fig.4 as a function of $\xi$.

$\frac{\eta}{\eta_{Cl}}$

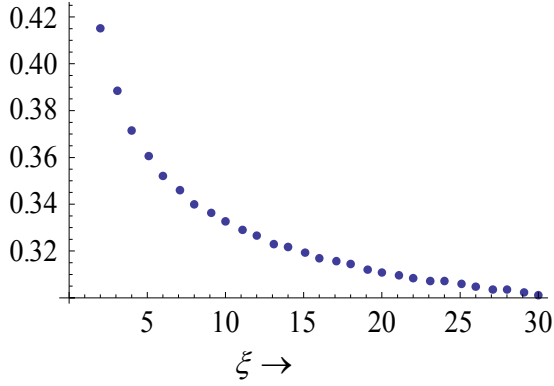

FIG. 4: (Color online) The hard sphere quantum viscosity as a function of $\xi$. The vertical axis is $\eta/\eta_{Cl}$, the ratio of the quantum result for the viscosity divided by the classical hard sphere result which is given by Eq. (10) with $\omega = 2$. At low values of $\xi$, the quantum calculation is the same as the classical result. At large $\xi$ the quantum result is ¼ the classical result.

**II.B.4 Quantum calculation of viscosity for a square well potential and the unitary limit**

Square well phase shifts are determined by the boundary conditions of continuity of the wave function and its slope at the square well radius. The phase shift $\delta_l$ is given by the equation $\tan\delta_l = \{kj_l'(x) - \gamma_l j_l(x)\}/\{kn_l'(x) - \gamma_l n_l(x)\}$ with $\gamma_l = \alpha j_l'(y)/j_l(y)$. The $\alpha = \sqrt{k^2 + |V_0|2\mu/\hbar^2}$, $y = \alpha R$, $x = kR$. The prime superscript on the spherical Bessel functions represent derivatives with respect to $x$ or $y$. The $y = n(E,V_0)x$ with $n(E,V_0) = \sqrt{1+|V_0|/E}$ the index of refraction of the classical description. Again, the low energy behavior will be considered as a baseline for comparison. In this limit the $S-D$ wave phase shift $\delta_2 - \delta_0 \approx -\delta_0$ and the bracket term in Eq. (16) is $2\sin^2\delta_0/3x^2$. The $S$–wave $\delta_0 = \arctan[(kR/\alpha R)\tan\alpha R - kR]$. An effective range approximation for $\delta_0$ reads $k\cot\delta_0 = -1/a_{sl} + r_0 k^2/2$. The scattering length $a_{sl} = R(1-\tan\alpha_0 R/\alpha_0 R)$ and the effective range is $r_0 = R - 1/\alpha_0^2 a_{sl} - R^3/3a_{sl}^2$. The $\alpha_0 = \sqrt{2\mu V_0/\hbar^2}$. For large $a_{sl}$, $r_0 \approx R$. A zero energy bound state appears when $\alpha_0 R = \pi/2$. Then $a_{sl} \to \infty$. Similarly, for a zero energy resonant like state $a_{sl} \to -\infty$. In an effective range approximation the $\omega$

$$\omega = 4\xi^4 \int_0^\infty dx \cdot e^{-\xi^2 x^2} x^7 \frac{1}{x^2} \left( \frac{(ka_{sl})^2}{1 + a_{sl}(a_{sl}-r_0)k^2 + (a_{sl}(a_{sl}-r_0)k^2)^2 \frac{r_0^2}{4(a_{sl}-r_0)^2}} \right) \quad (19)$$

As a further approximation, $r_0$ terms can be neglected when $|a_{sl}| \gg r_0$. Then

$$\omega = 4\frac{2}{3}\xi\left(\frac{2+\zeta(\zeta-1)-e^\zeta\zeta^3\Gamma(0,\zeta)}{2}\right) \tag{20}$$

where $\zeta = \xi R^2/a_{sl}^2 = (\lambda_T/a_{sl}\sqrt{2\pi})^2$ and $\Gamma(0,g)$ is a gamma function, with the special case $\Gamma(0,g) = E_1(g) = \int_g^\infty e^{-t}t^{-1}dt$. The limit $|a_{sl}| \to \infty$ is referred to as the unitary limit. In the unitary limit $\omega \to 8\xi/3 = 8(\lambda_T/R\sqrt{2\pi})^2/3$. Thus $\omega$ introduces quantum effects via the factor $\lambda_T$. The $S$-wave unitary or universal thermodynamic limit for $\eta$ is determined by the quantum wavelength $\lambda_T$ and is independent of the radius of the potential:

$$\eta \to \frac{15}{32}\frac{\sqrt{\pi k_B T m}}{\lambda_T^2} = \frac{15}{16}\frac{(\pi k_B T m)^{3/2}}{h^2} = \frac{15}{32}\sqrt{2\pi}\frac{\hbar}{\lambda_T^3}. \tag{21}$$

The last equality in Eq. (21) shows that the viscosity is proportional to Planck's constant divided by the quantum volume $\lambda_T^3$ and also related to a thermal momentum $p \sim \sqrt{k_B T m}$ as $\eta \sim p^3/h^2$. At $k_B T = 10 MeV$, the result of Eq.(21) is $1.03 \cdot 10^{-23} MeV \cdot s / fm^3$ and at $k_B T = 10 MeV$, it is $\eta = 0.032 \cdot 10^{-23} MeV \cdot s / fm^3$. By contrast the $S$-wave hard sphere limit can be written as $\eta = (5\sqrt{2\pi}/16)\cdot \hbar/(\lambda_T \pi R_C^2)$ but is $\hbar$ independent.

For identical particles, the $4\pi$ is changed to $8\pi$ in $\phi$ and the sum is over even or odd $l$ states. For particles with spin, spin factors appear. Identical spin $1/2$ fermions interacting through a $S$-wave, $l = 0$ state are coupled to a total spin zero singlet state. This introduces an additional factor of ¼ in $\phi$. The net effect is to reduce $\phi$ by ½, thereby increasing $\eta$ for fermions to twice the value given above to:

$$\eta = \frac{15}{16}\sqrt{2\pi}\frac{\hbar}{\lambda_T^3}. \tag{22}$$

The unitary limit for $\eta$ is independent of the potential used since it is based on an effective range result and with $a_{sl} \to \infty$. A calculation of $\eta$ with a delta shell potential [33] gave the same result and also the same result can be found in Ref. [26,27].

### II.B.5 The role of the nuclear tensor force and hard core

Some remarks on the role of the nuclear tensor force are as follows. The quantum theory of viscosity has $\eta \sim \Sigma(l+1)(l+2)/(2l+3)\cdot \sin^2(\delta_{l+2}-\delta_l)$. The nuclear non-central tensor force couples the $^3S_1$ phase shift to the $^3D_1$. In general, channels with spin $S = 1$ and angular momentum $L = J-1$ are coupled to channels with spin $S = 1$ and angular momentum $L = J + 1$. The analysis of phase shifts involves a parameter labeled $\varepsilon_J$ [43]. The centrifugal barrier suppresses the wave function at low energies where the phase shift $\delta_l \sim k^{2l+1}$ and at these energies $\varepsilon_J \sim k^{2l}$. In the isospin $I = 0$ channel ($np$ system) the nucleon-nucleon $^3S_1$ starts at the value $\pi$ because of the deuteron bound state and decrease rather quickly, reaching a value of $\sim \pi/2$ within the laboratory energy range of $5 MeV$. (See Fig.2-34, P264 in Ref. [43]). Thereafter the $^3S_1$ phase shift decreases only slightly over the energy range $5 MeV - 50 MeV$. Note that the low region near $\delta[^3S_1] \sim \pi$ does not contribute to $\eta$ since $\sin^2 \pi = 0$. The $^3D_1$ phase shift starts at a value equal to 0 and becomes somewhat significant at energies above $15 MeV$. Thus features related to the tensor force are suppressed for low enough temperatures. In the isospin $I = 1$ channel, the $nn, pp, np$ $S$-wave interaction is in the spin singlet $S = 0$ state or $^1S_0$ state. The tensor force does not act on a singlet state. In the $I = 1$ channel, the $nn, pp, np$ $P$-wave interaction is in the spin triplet $S = 1$ state which has $J = 0,1,2$ components or $^3P_{0,1,2}$ states. The tensor force couples a spin triplet $P$-wave to a spin triplet $l = 3$ $F$-wave where $J = 2,3,4$. Since $J$ must be the same, the tensor coupling is between $^3P_2$ and $^3F_2$. Again, the tensor force couples only becomes important at higher energies because the centrifugal barrier suppresses the $l = 3$ wave function compared to the $l = 1$ wave function.

Calculations of the effect of a hard core can be incorporated into a potential model. An analysis was represented in Ref. [29] for $S$-wave phase shifts arising from a square well interaction with a hard core. The $S$-wave phase shift for this potential is given b
$\delta_0 = \arctan[(kR_0/\alpha R)\tan\alpha(R_0 - R_C) - k(R_0 - R_C) - kR_C$. The $\alpha^2 = k^2 + \alpha_0^2$ and $\alpha_0 = \sqrt{2\mu|V_0|/\hbar^2}$. In an effective range approximation, the scattering length is $a_{sl} = R_0(1 - \tan\alpha_0(R_0 - R_C)/\alpha_0 R)$ and the effective range is now $r_{0+C} = r_0 + r_{0C}$ with $r_{0C} = R_C(1 - 2R_0/a_{sl} + R_0^2/a_{sl}^2 + 1/\alpha_0^2 a_{sl}^2)$. The hard core renormalizes $a_{sl}$ and $r_0$. The $R_C$ is the radius of the core and $R_0$ is the radius of the attractive square well. Results for higher partial waves where also given in Ref. [29].

The modifications due to a hard core in a classical approach are shown in the right figure in Fig. 1. Depending on the impact parameter $b$, index of refraction $n(E, V_0)$ and radius of the hard core $R_C$ the particle either misses the hard core or intersects it. For a given index of refraction the division between the two trajectories occurs at an impact parameter $b_m$ determined by $n \sin\theta_f = nR_C/R_0 = \sin\theta_i = b_m/R_0$ or simply $b_m = nR_C$. For $b > b_m$, the $\chi = -2(\theta_i - \theta_f)$ and the results of Eq. (12) can be used. If $b \leq b_m$, the $\chi$ is determined by $\chi = 2\theta_\beta - \pi - 2(\theta_i - \theta_f)$, $\sin\theta_\beta/R_0 = \sin(\pi - \theta_f)/R_C$. When $R_C \to R_0$, $\theta_\beta \to \pi - \theta_f$ and $\chi = \pi - 2\theta_i$, which is the hard sphere $\chi$. Subsection II.B.2 has $\phi, \omega \to 0$, and $\eta \to \infty$ as $n \to 1$. However with a hard core and with $n \approx 1$, the scattering is basically off the hard core since $b_m = nR_C \approx R_C$ and the results of subsection II.B.1 apply. For large $n$ such that $b_m = nR_C \geq R_0$ then all impact parameters lead to trajectories that intersect the hard core. For $n \to \infty$ all trajectories are directed toward the center inside the attractive region. Such trajectories then get reflected backwards against the incident path and leave the attractive region at the incident angle in the forward direction so that $\chi = \pi - 2\theta_i$. The result is again reflection off a hard sphere but now off a sphere of radius $R_0$. The ratio of the $\phi$ integrals at the limits $n \to 1$ and $n \to \infty$ is then $R_C^2 / R_0^2$.

### II.B.6 Low energy behavior of the viscosity of a dilute neutron gas

The viscosity of a dilute gas of neutrons will now be considered. A previous study of the second virial coefficient showed that the $S-$wave approximation accurately described the scattering up to temperatures of $\sim 15\ MeV$ before $P-$wave and $D-$wave contributions start to become significant. In a space symmetric $l = 0$ $S-$wave state, the neutrons are coupled to a total spin $S = 0$ antisymmetric state. In this channel the observed $S-$wave scattering length is $a_{sl} = -17.4\ fm$ and the effective range is $r_0 = 2.4\ fm$. A potential of the type considered in this paper that reproduces these properties has an attractive depth of $31.4 MeV$ with a radius of $2\ fm$ and hard core of radius $0.27\ fm$ [29]. The $\omega$ integral given by Eq. (20), when corrected for an effective range contribution, leads to a viscosity

$$\eta = \frac{(|a_{sl}| + r_0)}{|a_{sl}|} \left[ \frac{2}{2 + \zeta(\zeta - 1) - \zeta^3 e^\zeta \Gamma(0, \zeta)} \right] \left( \frac{15}{16} \sqrt{\pi}\ \frac{\hbar}{\lambda_T^3} \right) \qquad (23)$$

with $\zeta = \lambda_T^2 /(2\pi a_{sl}(a_{sl} - r_0)) = (\hbar c)^2 /(mc(a_{sl}(a_{sl} - r_0)k_B T) = 0.12 /(k_B T /1MeV)$. The factor $(a_{sl} - r_0)/a_{sl} = 19.8/17.4 = 1.138$. The last bracket term in Eq. 23 is the unitary limit. The middle square bracket term is equal to 0.9 at $k_B T = 0.5 MeV$, where $\zeta = .24$ and is equal to 0.99 at $k_B T = 10 MeV$, where $\zeta = 0.012$. Thus the viscosity of a neutron gas is within $\sim 10\%$ of its unitary limit when $k_B T \geq 0.5 MeV$. In the above equation the factor in square bracket is very accurately approximated by $1 + \zeta/3$ for the entire range of $\zeta$. The $\zeta/3$ is obtained from the asymptotic value of $2 + \zeta(\zeta - 1) - \zeta^3 e^\zeta \Gamma(0, \zeta)$ for large $\zeta$. The value $1$ is the unitary limit. Thus the viscosity is very accurately given by

$$\eta = \left( \frac{(|a_{sl}| + r_0)}{|a_{sl}|} + \frac{\lambda_T^2}{3 \cdot 2\pi a_{sl}^2} \right)\left( \frac{15}{16} \sqrt{2\pi}\ \frac{\hbar}{\lambda_T^3} \right). \qquad (24)$$

The first term alone in the first bracket is larger than the unitary limit since $(|a_{sl}| + r_0)/|a_{sl}| > 1$. The second term, with an overall $1/(a_{sl}^2 \lambda_T)$ dependence, becomes large when $\lambda_T /(\sqrt{2\pi}|a_{sl}|) \sim \sqrt{3}$. For $|a_{sl}| = 17.4\ fm$, the $\lambda_T \sim 76\ fm$ and thus $T \sim 1/20 MeV$ for the second term to become comparable to the first term.

## II. Entropy and viscosity to entropy density ratio

### III.A.1 General considerations

The equation of state $EOS$ can be used to obtain the interaction part of the entropy using the Maxwell relation $(\partial P / \partial T)_V = (\partial S / \partial V)_T$. The $EOS$ to second order in the virial expansion is $P = k_B T (A/V - \hat{b}_2 A^2 / V^2)$. The energy including the interaction energy [29] obtained from the thermodynamic identity $(\partial E / \partial V)_T = T(\partial P / \partial V)_T - P$ leads to $E(V,T) = (3/2) A k_B T + k_B T^2 (d\hat{b}_2 / dT)$. The temperature dependence of $E(V,T)$ and $S(V,T)$ are connected by $(\partial E / \partial T)_V = T(\partial S / \partial T)_V$. Also $(\partial E / \partial V)_T = T(\partial S / \partial V)_T - P$. The $\hat{b}_2 = \hat{b}_{2,exc} + \hat{b}_{2,int}$ where $\hat{b}_{2,exc}$ is the exchange part of the virial coefficient and $\hat{b}_{2,int}$ arises from interactions. The $\hat{b}_{2,exc} = \pm (1/2^{7/2}) \lambda_T^3 / g_S$ with $g_S = 2S+1$ and the + sign is for bosons and the minus sign is for fermions. The interaction $\hat{b}_{2,int}$ is

$$\frac{\hat{b}_{2,int}}{\lambda_T^3 2^{3/2} / g_S^2} = \sum_{E_b(J)} g_J \exp(-\frac{E_b(J)}{k_B T}) + \frac{1}{\pi} \sum_J g_J \int \frac{d\delta_J}{dk} \exp(-bk^2) dk \equiv B_b + B_C. \quad (25)$$

The $b = \lambda_T^2 / 2\pi$. The $B_b$ is the bound part of $\hat{b}_{2,int}$ with $E_b(J)$ the energy of a bound state with spin $J$ and degeneracy $g_J$. The second term involving $d\delta_l / dk$ is a term due to Beth and Uhlenbeck [40-42] and it reflects continuum correlations and is labeled $B_C$. The interaction part of the entropy is obtained from $\hat{b}_{2,int}$ using $S_{int} = k_B (A^2 / V) d(T\hat{b}_{2,int}) / dT$. The interaction entropy can also be written in terms of $\xi = b / R_C^2 = \lambda_T^2 / 2\pi R_C^2$ as a variable which is useful when evaluating $S_{int}$ for a hard core potential. Specifically,

$$S_{int} = -k_B \frac{A^2}{V} \frac{2^{3/2}}{g_S^2} (2\pi R_C^2)^{3/2} \left( \xi^2 \frac{d}{d\xi} \xi^{1/2} B_C(\xi) \right). \quad (26)$$

A similar expression applies in an effective range approximation with $\xi$ replaced with $\zeta == \lambda_T^2 / (2\pi (a_{sl}(a_{sl} - r_O)))$ and $R_C^2$ replaced by $a_{sl}(a_{sl} - r_O)$.

### III.A.2 S-wave effective range results and the unitary limit

The $S$-wave effective range approximation leads to [29]

$$\frac{d\delta_0}{dk} = -\frac{a_{sl}}{1 + a_{sl}(a_{sl} - r_0)k^2 + (r_0 a_{sl})^2 k^4 / 4} (1 + \frac{r_0 a_{sl}}{2} k^2) \approx$$

$$-\frac{a_{sl}}{1 + a_{sl}(a_{sl} - r_0)k^2} (1 + \frac{r_0 a_{sl}}{2} k^2) . \quad (27)$$

The last approximation in Eq. (27) omits the $k^4$ term which turns out to accurately describe the behavior of $B_C$ unless $a_{sl} \sim r_0$. Also if $r_0 > a_{sl} > 0$ the $k^4$ is necessary to keep the integral from diverging. Away from these regions, the $B_C$ is given by

$$B_C = -\frac{a_{sl}(2a_{sl}^2 - 3r_0 a_{sl})}{4(a_{sl}^2 - a_{sl}r_0)^{3/2}} \exp(\frac{b}{a_{sl}^2 - a_{sl}r_0}) Erfc(\sqrt{\frac{b}{(a_{sl}^2 - a_{sl}r_0)}}) - \frac{a_{sl}^2 r_0}{4(a_{sl}^2 - a_{sl}r_0)\sqrt{\pi b}}. \quad (28)$$

The unitary limit is $a_{sl} \to \pm\infty$. In this limit, the continuum contribution $B_C$ to $\hat{b}_{int}$ is

$$B_C = -sign[a_{sl}] \cdot \frac{1}{2} - \frac{1}{4\sqrt{\pi}} \frac{r_0}{\sqrt{b}} + \frac{1}{32\sqrt{\pi}} (\frac{r_0}{\sqrt{b}})^3 = \frac{1}{2} - \frac{\sqrt{2}}{4} \frac{r_0}{\lambda_T} + \frac{3\sqrt{2}\pi}{16} (\frac{r_0}{\lambda_T})^3. \quad (29)$$

When $a_{sl} \to +\infty$, the bound state term of 1 and continuum term $-sign[a_{sl}]/2 = -1/2$ give $+\frac{1}{2}$ which is the same result for a zero energy resonance with $g_J = 1$. The $r_0/\lambda_T$ part of $B_C$ does not contribute to the $S_{int}$. The $S_{int}$ in the limit $a_{sl} \to -\infty$ is

$$S_{int} = -k_B \frac{A^2}{V} \lambda_T^3 \frac{2^{3/2}}{g_S^2} \left( \frac{1}{4} - \frac{3\sqrt{2}\pi}{16} \frac{r_0^3}{\lambda_T^3} \right). \quad (30)$$

Even in the limit of infinite scattering length, features related to the effective range persist [29]. For $S_{int}$ the residual effects of the effective range have a cubic dependence.

### III.A.3 Entropy for a neutron gas

For spin ½ fermions, and in particular for a pair of neutrons, the $S$-wave interaction is in the spin singlet $S = 0$ state while the $P$-wave interaction is in the spin triplet $S = 1$ state and has $J = 0,1,2$. The $J$-weighted average of the $^3P_0, ^3P_1, ^3P_2$ phase shifts given by $\bar{\delta}(^3P) = \{5\delta(^3P_2) + 3\delta(^3P_1) + 1\delta(^3P_0)\}/9$ is small [43]. Both a nuclear spin-orbit force and tensor force are necessary to explain features associated with the $P$-wave phase shifts. A spin-orbit force given by $V_{LS}(r)\vec{L} \cdot \vec{S} = V_{LS}(r)(J(J+1) - l(l+1) - s(s+1))/2$ cannot account for the behavior of the triplet $P$-wave phase shifts. If the spin-orbit were the only spin dependent force the $^3P_1$ $J = 1$ phase shift would be intermediate between the $^3P_2, J = 2$ and $^3P_0, J = 0$ phase shifts from the $J(J+1)$ dependence in $\vec{L} \cdot \vec{S}$. This feature is not consistent with experimental results shown in Fig. 5. The $D$-wave interaction is in the spin singlet $S = 0, J = 2$ state and has $\delta(^1D_2)$. The $\delta(^1D_2)$ is small as shown in Fig. 5. Thus the neutron gas entropy is mainly dominated by the $S$-wave term. In an effective range theory the $S$-wave entropy is given by Eq. (30).

.

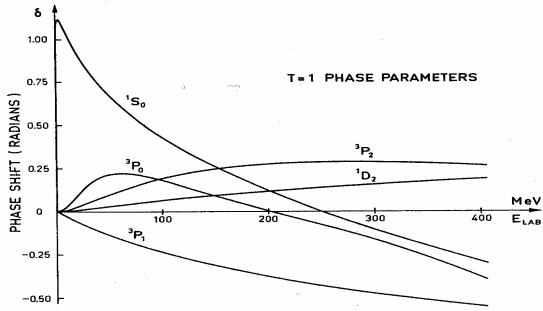

FIG. 5. The nucleon-nucleon isospin T=1 phase shifts in radians versus $E_{Lab}(MeV)$. The figure appears in part in Bohr and Mottleson [43].

### III.A.4 Entropy of a hard sphere gas

For a hard sphere gas the bound state contribution $B_b = 0$ and $B_C$ is obtained from $\tan \delta_l = j_l(x)/\eta_l(x)$, $x = kR_C$ and $d\delta_l/dx = -1/(x^2(j_l^2(x) + \eta_l^2(x)))$. Appendix A gives a complete discussion of $S_{int}$ and also $\eta$. Using $\xi = b/R_C^2 = \lambda_T^2/2\pi R_C^2$, the $S$–wave contribution to $B_C = -\xi^{-1/2}/2$ and the interaction entropy $S_{int} = 0$ by Eq. (26).
At very high energy, or as $x \to \infty$, the following scaling relation [29] was noted:

$$S_2 \equiv \sum_{l=0}^{x}(2l+1)\frac{1}{(j_l^2(x)+\eta_l^2(x))} \to \frac{2}{3}x^4. \tag{31}$$

In the limit that many terms contribute to $B_C$

$$B_C = \frac{1}{\pi}\int dk \sum_{l=0}^{\infty}(2l+1)\frac{d\delta_l}{dk}\exp(-bk^2) = -\frac{\sqrt{2}}{3}\pi\frac{R_C^3}{\lambda_T^3}. \tag{32}$$

For spin zero bosons and spin ½ fermions these scaling or geometric limits are reduced by ½ -see appendix A. For high $T$ the interaction entropy is then

$$S_{int} \to -\frac{1}{2}(\frac{4}{3}\pi R_C^3)k_B\frac{A^2}{V}. \tag{33}$$

The value $S_{int} = -V_C k_B A^2/2V$, with hard core volume $V_C = 4\pi R_C^3/3$, is the semi classical limit of the interaction entropy. The result can be obtained from the quantum result by taking $\hbar \to 0$ or $\xi \to 0$. The $S_{int}$ is density $A/V$ dependent but temperature $T$ independent. This result will be compared with a van der Waals gas in the next subsection.
For a monatomic ideal gas of $A$ nucleons, the entropy is given by the Sakur-Tetrode expression [28] which is $S = S_{id} = Ak_B \ln(e^{5/2}Vg_S/A\lambda_T^3)$.

$$S = Ak_B \left( \frac{5}{2} + \ln \frac{Vg_S}{A\lambda_T^3} + \{\pm \frac{1}{2^{7/2}} \frac{A\lambda_T^3}{Vg_S} + 2(2^{-4} - 3^{-5/2})(\frac{A\lambda_T^3}{g_S V})^2\} + \frac{A}{V} \frac{d}{dT}(T\hat{b}_{2,\text{int}}) \right) \ldots \quad (34)$$

with the $+$ sign for fermions and the $-$ sign for bosons in the $\pm$ signs in $S$.

**II.A.5. Comparison with a van der Waals gas; comments on the role of a liquid/gas phase transition, critical point fluctuations and critical opalescence.**

The *EOS* of a van der Waal gas is $(P + a(T)(A^2/V^2))(V - b_{ex}A) = Ak_B T$. The $b_{ex}$ is the excluded volume term and the $a(T)$ arises from two particle attractive interactions which can be temperature dependent. Using the Maxwell identity listed above, the entropy is $S = Ak_B \ln(V - b_{ex}A)T^{3/2} + (A^2/V)(da(T)/dT) + C_S$, with $C_S$ a constant. Spin entropy will be omitted. The Sakur-Tetrode law for the entropy is a quantum theory result for an ideal gas (no $b, a(T)$) and gives $C_S = -Ak_B \ln(Ah^3/(2\pi mk)^{3/2}) + (5/2)Ak_B$. A constant $(5/2)Ak_B$ also appears in the Sakur-Tetrode law from $E + PV$. The entropy is then

$$S = Ak_B \ln e^{5/2} \frac{(V - b_{ex}A)}{A\lambda_T^3} + \frac{da(T)}{dT} \frac{A^2}{V}. \quad (35)$$

In the dilute limit the $V \gg b_{ex}A$. Expanding the logarithm gives

$$S \approx Ak_B \ln e^{5/2} \frac{V}{\lambda_T^3} - k_B A \frac{A}{V} b_{ex} + \frac{da(T)}{dT} \frac{A^2}{V}. \quad (36)$$

The $-k_B A^2 b_{ex}/V$ factor is the same as the scaling limit of the hard sphere quantum gas apart from a factor of 2 reduction for identical particles.

One of the interesting features associated with the van der Waals EOS is its connection with a liquid gas phase transition when combined with a Maxwell construction which is introduced to eliminate regions of negative isothermal compressibility. A review of the nuclear liquid gas phase transition can be found in Ref. [44]. The van der Waal model parallels a density functional approach based on a Skyrme interaction [45,46]. The behavior of viscosity with temperature in a liquid is considerable different than in a gas. The viscosity of a liquid decreases rather rapidly with $T$ while that of a gas increases with $T$ [30,31]. As an example of the rapid decrease with $T$ is the viscosity of water which decreases by a factor of about 6 from its freezing point $273^0 K$ to its boiling point $373^0 K$. The result of Eq. (1,2) gives a slow increase of $\eta$ as $\sqrt{T}$ for a gas from $\hat{v}$, the mean speed of the particles. The ratio for $\eta(373^0 K)/\eta(273^0 K) = \sqrt{373/273} = 1.17$. In the unitary limit, $\eta \sim T^{3/2}$ and $\eta(373^0 K)/\eta(273^0 K) = 1.17^3 = 1.6$.

A second interesting feature is the presence of a critical point where large fluctuations in density occur. The phenomena of critical opalescence is a characteristic feature of these critical point density fluctuations where the scattering cross section increases dramatically because droplets of all sizes are present. A simple model is the Fisher model [47] of a critical point where the distribution of cluster sizes falls as a scale invariant power law. The number of clusters $n_k$ of size $k$ varies as $n_k \sim 1/k^\tau$ with $\tau$ a critical exponent. The present work evaluates the viscosity of one type of particle with a single fixed size. The viscosity of a given type of cluster depends on mass $m$ and radius $R = D/2$ of a cluster as $\sqrt{m}/D^2 \sim 1/R^{1/2}$ using the simple expression of Eq. 2. The radius term comes from the cross section through the mean free path and mass term is present from speed factors. The mass of a cluster with $A$ nucleons varies as $m \sim m_p A \sim R^3$. At a critical point, the matter is now made of many different types of clusters, but subject to a constraint of overall mass conservation which reads $A = \Sigma_k k n_k$. Because of clustering, the multiplicity $m = \Sigma_k n_k$ is reduced from $A$ to a much smaller number. Scattering can now also occur between particles of different sizes.

In a ratio $\eta/s$ the entropy must be considered when bound states are formed. The entropy is greatly affected by large changes in the multiplicity. To see this feature consider the factor $(5/2)k_B$ that appears in the entropy. Each particle (monomer, dimer,…) adds $3/2 k_b T$ to the energy and $k_B T$ to $PV$. Since $TS = E + PV - \Sigma \mu_k n_k$ for a mixture of non-interacting ideal gases, the entropy $S = 5 m k_B T/2 - \sum_k \mu_k n_k$.

A second way to explore the behavior of a system around a critical behavior is based on an order parameter expansion. The simplest order parameter theory is a Ginzburg-Landau mean field approach which has been used to study the fluctuations in density near the critical point in nuclear systems [48]. Large density fluctuations lead to a divergence of the isothermal compressibility with an associated exponent describing the divergence. Similarly, fluctuations in the energy determine the heat capacity. The viscosity is determined by fluctuations in the stress tensor. Each divergence has a critical exponent. The critical exponents of mean field theories are not those observed experimentally and improved techniques based on renormalization group methods have been developed [41]. The divergence in the viscosity arises from the coupling of the transverse velocity to the order parameter density fluctuations. Past calculations [49] show that the viscosity diverges with an exponent $8/15 \pi^2$ and somewhat larger values have been noted from dynamic renormalization group techniques [50].

**III.A.6. Isothermal compressibility of a dilute neutron gas in the unitary limit .**

As already noted, the behavior of the isothermal compressibility $\kappa_T$ around a scale free critical point is an important thermodynamic quantity in such studies. The behavior of $\kappa_T$ in the unitary limit also has interesting features. In the dilute gas limit these features can be obtained from the second virial coefficient through the equation

$$\kappa_T = -\frac{1}{V}\frac{\partial V}{\partial P}\bigg|_T = \frac{1}{A k_B T/V - 2\hat{b}_2 A^2 k_B T/V^2} = \frac{\kappa_{T,ideal}}{1 - 2\hat{b}_2 A/V} \quad . \tag{37}$$

The $\kappa_{T,ideal} = 1/((A/V)k_B T)$ is the ideal gas compressibility. The $\kappa_T$ should be positive for the mechanical stability of a gas. For an ideal Bose gas the isothermal compressibility becomes infinite at the condensation point with the singularity arising from the sum of an infinite series of terms in a virial expansion. For an ideal gas these terms arise solely from symmetrization terms of the type shown in {} brackets in Eq. (34). For an imperfect Bose gas this divergence is removed [41]. For fermions, if $\hat{b}_2$ is positive, interaction effects are more important than fermionic antisymmetrization effects and $\kappa_T$ will have a peak as $T$ increases from low to high temperatures as discussed in Ref. [51]. The presence of a hard core potential or strong repulsive three body terms is also important in understanding the compressibility of nuclear matter. Repulsive components are necessary for saturation of cold nuclear matter at the proper density.

The unitary limit for a neutron gas is $\hat{b}_2 = \lambda_T^3(-1/2^{9/2} + 2^{3/2}/(2^2 \cdot 2))$ when effective range corrections are neglected. Then $B_C = 1/2$ and the additional factor $-1/2^{9/2}$ is from antisymmetrization of the pair of neutrons while the $1/2^2 = 1/g_S^2$. The compressibility in the dilute neutron gas in the unitary limit is:

$$\kappa_T = \xi_\kappa \kappa_{T,ideal}, \quad \xi_\kappa = \frac{1}{1 - 2(\frac{7}{2^{9/2}})\frac{A}{V}\lambda_T^3} = \frac{1}{1 - 2(\frac{7}{2^{9/2}})z} \quad . \tag{38}$$

When effective range corrections are neglected the isothermal compressibility takes on this very simple form. Effective range corrections to the compressibility can be included using Eq. (27). Since these corrections involve $B_C$ they contain both linear and cubic terms in the ratio $r_0/\lambda_T$ and reference to interaction potential appears in $\kappa_T$. A neutron gas is mechanically stable when $\kappa_T > 0$ and this condition is realized for low $z$. A similar result for bosons has $1/\xi_\kappa = 1 - 2 \cdot (17/2^{7/2})z$. For a hard sphere $S$-wave Fermi gas the $\kappa_T = \kappa_{T,ideal}/(1 + z/2^{7/2} + (2z/g_S^2)(2R_C/\lambda_T))$ is positive.

**III.B Viscosity to entropy density ratio $\eta/s$**

Low viscosity to entropy density is associated with a nearly perfect fluid [4,5]. How perfect is a gas of nucleons? How close is a nuclear system to the AdS/CFT string theory minimum $\eta/s = (1/4\pi)\hbar/k_B$ [9], where $s = S/V$? As a first step in trying to answer these questions, a simple one component system of spin ½ fermions will be considered and with little additional effort results for spin 0 bosons will be given. This study is done in a unitary limit in the next subsection B.1 neglecting a small correction to the entropy density in the unitary limit from the effective range. In subsection B.2 and B.3, the $\eta/s$ ratio is studied in a system not at or close to the unitary limit of infinite scattering length. In particular the effective range theory and also the hard sphere gas are used since simple analytic results can be obtained. The main difference with the unitary limit is that a dimensionless variable (labeled $y$) involving either the scattering length or the radius of the hard core to the interparticle spacing appears in $\eta/s$. The $\eta/s$ becomes a function of $y$ and the fugacity $z = \rho \lambda_T^3$.

### III.B.1 $\eta/s$ in fermionic and bosonic systems in the $S$-wave unitary limit $a_{sl} \to -\infty$.

As a first example only $S$-wave interactions in the unitary limit will be taken. The spin ½ fermion case is approximately realized in pure neutron matter where the neutron pairs are coupled to total spin $S = 0$ and have a large negative scattering length. The fermionic case is compared to the bosonic case for spin zero bosons. Two main differences arise between the fermionic case and the bosonic case. The first difference is the change to symmetrization for bosons from antisymmetrization for fermions. The second main difference arises from spin degeneracy factor $g_S$ and associated spin entropy. The entropy density to lowest order in antisymmetrization corrections for fermions is

$$s = \rho k_B \left( \frac{5}{2} + \ln g_S - \ln \rho \lambda_T^3 \pm \frac{1}{2^{7/2} g_S} \rho \lambda_T^3 - \frac{2^{3/2}}{(g_S)^2} \frac{1}{2 \cdot 2} \rho \lambda_T^3 \right). \tag{39}$$

The $\eta = g_S^2 15\sqrt{2}\pi \hbar /(\lambda_T^3 64)$ for $S$-wave fermions or bosons in the unitary limit. Defining the standard thermodynamic variable called the fugacity as $z = \rho \lambda_T^3 = e^{\mu/k_B T}$, the $\eta/s$ is

$$\left. \frac{\eta}{s} \right|_j = g_S^2 \frac{15}{64} \sqrt{2}\pi \frac{\hbar}{k_B} \frac{1}{z} \frac{1}{\frac{5}{2} + \ln g_S - \ln z + ((-1)^{g_S} \frac{1}{2^{7/2} g_S} - \frac{2^{3/2}}{g_S^2} \frac{1}{4})z} \tag{40}$$

The minimum of $\eta/s$ or maximum of $s/\eta$ occurs at $z_m$ given by

$$3/2 + \ln g_S - \ln z + 2((-1)^{g_S} \frac{1}{2^{7/2} g_S} - \frac{2^{3/2}}{g_S^2} \frac{1}{4})z = 0 \tag{41}$$

which has a solution $z_{m,f} = 3.52$ for fermions and $z_{m,b} = 0.965$ for bosons. For fermions the value $z_{m,f} = 3.52$ is in a region of $z$ where higher order terms must be included. The problem of a large value of $z_m$ for fermions resides more in the higher order interaction effects from three body, four, …body terms rather than higher order antisymmetrization effects. The $\eta/s$ at $z_{m,f} = 3.52$ is

$$\left. \frac{\eta}{s} \right|_f = .806 \frac{\hbar}{k_B} = 10.14 \frac{1}{4\pi} \frac{\hbar}{k_B} \tag{42}$$

For bosons $z_{m,b} = 0.965$ and

$$\left.\frac{\eta}{s}\right|_b = .610 \frac{\hbar}{k_B} = 7.67 \frac{1}{4\pi} \frac{\hbar}{k_B} \tag{43}$$

The $z_{m,f} = 3.52$ and $z_{m,b} = 0.965$ are in a regions of negative isothermal compressibility where the system is mechanically unstable. Higher order effects beyond the second virial coefficient should be included for a proper description at these high values of the fugacity. The next two subsections look at the question of $\eta/s$ away from the unitary limit.

**B.2 $\eta, s,$ & $\eta/s$ away from the unitary limit; effective range approximation**

The behavior of viscosity, entropy density and their ratio away from the unitary limit $a_{sl} \to -\infty$ is developed in this subsection. Again, the $k^4$ correction in the denominator of Eq. [27] will be neglected which imposes restrictions on the relation of $a_{sl}$ to $r_0$ as already discussed. The region $a_{sl} > 0$ contains at least one bound state and scattering off bound states must be included as well as corrections to the entropy from $k^4$ corrections. Therefore, the case $a_{sl} < 0$ will only be developed for now. To keep the results simple the effective range $r_0$ corrections will also be omitted. Only $S$-wave terms will be considered in this section which requires $k_B T \leq 25 MeV$ or $\lambda_T \geq 3 fm$ for a hadronic system. The viscosity for a fermionic of bosonic system is then given by

$$\eta = \left(1 + \frac{\lambda_T^2}{3 \cdot 2\pi a_{sl}^2}\right)\left(g_S^2 \frac{15}{64}\sqrt{2\pi}\frac{\hbar}{\lambda_T^3}\right). \tag{44}$$

For the same effective range approximation the interaction entropy is

$$S_{int} = -k_B A \frac{A}{V} \frac{2^{3/2}}{g_S^2} \frac{(2\pi)^{3/2} 2|a_{sl}^3|}{4} f_S(\zeta) \tag{45}$$

with $f_S(\zeta) = \zeta^2[1/(2\zeta^{1/2})e^\zeta erfc(\sqrt{\zeta}) + \zeta^{1/2}e^\zeta erfc(\sqrt{\zeta}) - 1/\sqrt{\pi}]$. In the unitary limit $a_{sl} \to -\infty, \zeta \to 0, f_S(\zeta) \to \zeta^{3/2}/2 = (\lambda_T^3/(2\pi a_{sl}^2)^{3/2}/2$ and the expression of Eq. (27) results. When $\zeta \gg 1$ then $f_S(\zeta) \to 1/(2\sqrt{\pi})$ and $S_{int}$ depends on the interaction potential through the factor $|a_{sl}^3|$. The $S_{int}$ is added to the ideal gas entropy to give $S = k_B A[5/2 - \ln(A\lambda_T^3/V) - (1/2^{7/2})A\lambda_T^3/V] + S_{int}$. The $\eta/s$ then follows and is

$$\frac{\eta}{s} = \frac{\hbar}{k_B}\left(g_S^2 \frac{15\sqrt{2}}{64}\pi\right)\frac{F_\eta(z,y)}{z(5/2 - \ln z + (-1)^{g_S} 2^{-7/2} z/g_S - F_S(z,y))} \tag{46}$$

with $F_\eta(z,y) = 1 + z^{2/3}/(3y^2)$ and $F_S(z,y) = (2^{3/2}/g_S^2)\cdot(2y^3/4)f_S(z^{2/3}/y^2)$. The $f_S(z^{2/3}/y^2) = f_S(\zeta)$ with $\zeta = z^{2/3}/y^2 = \lambda_T^3/(2\pi a_{sl}^2)$. A dimensionless variable $y$ given by

$$y = \sqrt{2\pi}|a_{sl}|/(V/A)^{1/3} \tag{47}$$

is introduced besides the fugacity variable $z = (A/V)\lambda_T^3$ which is a thermodynamic variable. The $y$ variable is a measure of the ratio of the scattering length to interparticle spacing. Including effective range corrections adds another variable which can be defined in a manner similar to $y$ with $|a_{sl}|$ replaced by $r_0$; namely $x = \sqrt{2\pi}r_0/(V/A)^{1/3}$ which can be viewed as a measure of diluteness since $r_0 \approx R_0$ the radius of the well $\approx$ diameter of the particle for a short range nuclear potential. A plot of $\eta/s$ versus $z$ for various $y$ for fermions is shown in Fig. 6. Again, the minima occur in a region of negative isothermal compressibility and large fugacity.

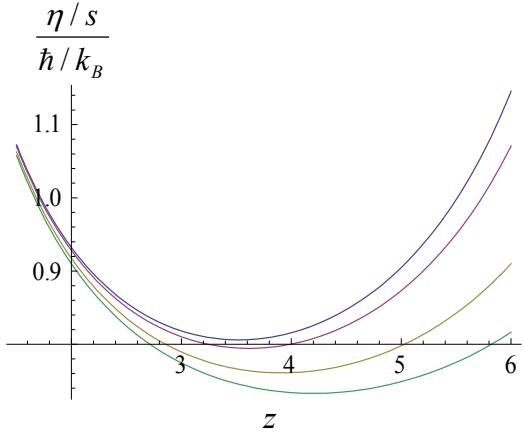

FIG. 6. (Color online) The ratio $\eta/s$ in units of $\hbar/k_B$ versus $z$ for various $y$ for fermions. The values of $y$ are $y = \infty, 100, 25, 15$. The lowest curve has the lowest $y$ and higher curves have increasing $y$. Taking a minimum of $\eta/s \sim 0.75\,\hbar/k_B$, then $\eta/s \sim 9.4(\hbar/(4\pi k_B))$.

### III.B.2 $\eta/s$ for a hard sphere gas; $S$-wave model

Some simple models of viscosity treat the collision between particles as a hard sphere collisions. This subsection explores properties of such a gas, and in particular the $\eta/s$ ratio, in a quantum description. The problem of negative isothermal compressibility does not arise as shown in III.A.6 for fermions. Only $S$-wave interactions are considered in this subsection. Results for all partial waves are given in the appendix. Results are shown in Fig. 7.

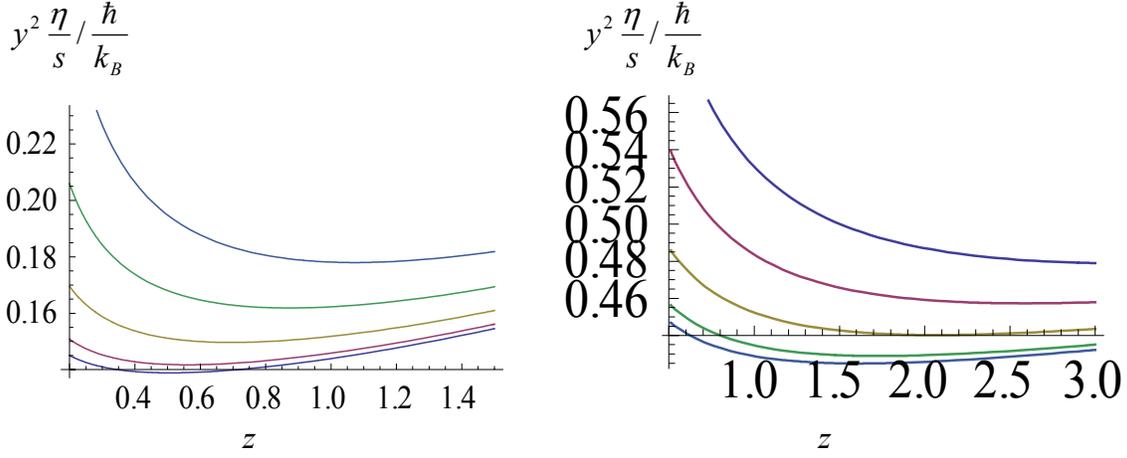

FIG. 7. (Color online) The ratio $\eta/s$ versus fugacity $z$ for various $y$ for a hard sphere S-wave interaction. The vertical axis is $y^2\eta/s$ divided by $\hbar/k_B$. The $y = \sqrt{2\pi}|a_{sl}|/(V/A)^{1/3}$. The left figure is the Bose gas and the right figure is the Fermi gas. The lowest curve in each figure has $y = 0$ and each successive curve has 0.1 unit higher in $y$ with the uppermost curve at $y = 0.4$. The minima in $\eta/s$ shift to higher $z$ with increasing $y$. The vertical axis does not have a large spread in values since the curves are rather flat expect for the $y = 0.4$ curve. A comparison of the $S$–wave hard sphere model with the geometric limit is given in appendix A.

The $\omega$, including an extra factor of 2 for identical fermions or bosons and a factor $1/g_S^2 = \frac{1}{4}$ for spin ½ fermions coupled to spin 0, is

$$\omega = \frac{2}{g_S^2} 4\xi^4 \frac{2}{3} \frac{1}{8\xi^{11/2}} (\sqrt{\xi}(9\xi - 2) + \{4 + 5\xi(3\xi - 4)\} \cdot D(\frac{1}{\sqrt{\xi}})) \qquad (48)$$

where $D(1/\sqrt{\xi})$ is the Dawson F function which has an expansion for large $\sqrt{\xi}$ given by the $D(\xi) = \xi^{-1/2} \Sigma_{n=0} (-1)^n 2^n /(\xi^n (2n-1)!!)$. The ratio $\eta/s$ is then given by the equation

$$\frac{\eta}{s} = \frac{\hbar}{k_B} \frac{5}{8} \sqrt{2\pi} \frac{2}{y^2} \frac{1}{z^{1/3}} \frac{1}{\omega(z^{2/3}/y^2)} \frac{1}{(5/2 + \ln g_S - \ln z \pm z/(2^{7/2} g_S))} . \qquad (49)$$

The $\xi = z^{2/3}/y^2$.

## IV Viscosity and collective flow in nuclear collisions
### 4.1 Kinetic flow in a relaxation approach to the Boltzmann equation

The kinetic flow tensor is $F_{ij} = \Sigma_{k=1}^{N}(P_i(k)P_j(k)/2m(k))$ where $i, j$ are components of the momentum $\vec{P}(k)$ of particle $k$ which has mass $m(k) = m$. The sum is over all particles. The following ansatz will be used for the phase space distribution for a system of two symmetric colliding nuclei:

$$f^{(0)} = n(\vec{r},t)\left[\frac{m}{2\pi\theta(\vec{r},t)}\right]^{1/2}\left(\exp(-\frac{m(\vec{v}-\vec{u}(\vec{r},t))^2}{2\theta(\vec{r},t)}) + \exp(-\frac{m(\vec{v}+\vec{u}(\vec{r},t))^2}{2\theta(\vec{r},t)})\right), \quad (50)$$

The $f^{(0)} = f^{(0)}(\vec{r},\vec{v},t)$ is taken as the zero'th order approximation to the Boltzmann equation which has a phase space density $f(\vec{r},\vec{v},t) = f$. The $\vec{u}(\vec{r},t) \equiv \vec{u} = <\vec{v}>$ and $3\theta(\vec{r},t)/2 = m<|\vec{v}-\vec{u}(\vec{r},t)|^2>/2$ where the expectation values are taken with $f^{(0)}$. The $n(\vec{r},t) = \int d^3v f(\vec{r},\vec{v},t)$. The Boltzmann equation is $(\partial/\partial t + \vec{v}\cdot\nabla_r + (\vec{F}/m)\cdot\nabla_v)f = (\partial f/\partial t)_{coll}$. The $\vec{F}$ is a force term from an external field or Hartree-Fock field. In a relaxation time approximation the collision term $[\partial f/\partial t]_{coll} = -(f - f^{(0)})/\tau_R$. The correction $f - f^{(0)} = g$ is taken to be small and the left hand side of the Boltzmann equation is evaluated by taking $f = f^{(0)}$ giving

$$g = -\tau_R\left[\frac{1}{\theta}\frac{\partial\theta}{\partial x_i}U_i\left(\frac{m}{2\theta}U^2 - \frac{5}{2}\right) + \frac{1}{\theta}\Lambda_{ij}\left(U_iU_j - \frac{1}{3}\delta_{ij}U^2\right)\right]f^{(0)}. \quad (51)$$

Repeated indices are summed over. The $\Lambda_{ij} = m/2(\partial u_i/\partial x_j + \partial u_j/\partial x_i)$ and $\vec{U} = \vec{v} - \vec{u}$. The pressure tensor $\vec{P}$ has elements given by $P_{ij} = mn(\vec{r},t)\langle U_iU_j\rangle = P\delta_{ij} + P'_{ij} = \rho\theta\delta_{ij} - (2\mu/m)(\Lambda_{ij} - (m/3)\delta_{ij}\vec{\nabla}\cdot\vec{u})$ with $\mu = \rho\tau_R\theta$ and $\rho$ is the number density. An important connection is the relation of the kinetic flow tensor to the pressure tensor. This result is $F_{ij} = (m/2)\int n(\vec{r},t)u_i(\vec{r},t)u_j(\vec{r},t)d^3r + \int P_{ij}d^3r \equiv F_{ij}(u) + F_{ij}(P)$. The off diagonal part of $F_{ij}$ has a term coming from $F_{ij}(u)$ which is the collective flow term and $F_{ij}(P')$ which is related to the shear viscosity through the relationship

$$F_{ij}(P') = \int P'_{ij}d^3r = -(\mu/m)\Lambda_{ij}\Omega_V. \quad (52)$$

The $\Omega_V$ is the volume of the system. The minus sign in $F_{ij}(P')$ shows that the shear viscosity cancels part of the collective flow. Early calculations were presented in Ref. [1] and the results showed that the cancellation could be significant.

### 4.1 Viscosity and Reynolds number; laminar or turbulent flow?

The Reynolds number $R_Y$ is defined as $R_Y = (d\rho mu/\eta)$ where $d$ is a characteristic length.

The quantity $\eta/m\rho \equiv \nu_\eta$ is the kinematic viscosity so that $R_Y = du/\nu_\eta$. The connection with the string theory limit on $\eta/s$ arises from the connection of the entropy density to $\rho$ since $s \sim k_B \rho$. Thus $R_Y \sim dm/k_B \eta/s$. As a first approximation the simple expression $\eta = (1/3)nm\hat{v}l_\lambda$ can be used to give $R_Y$ as $R_Y = 3(u/\hat{v}) \cdot (d/l_\lambda)$. In a collision the characteristic distance $d$ is of the order of the size of the nucleus. The collective velocity $u$ is of the order of the incident velocity in a medium energy collision. The thermal speed is $\hat{v}/c = \sqrt{8k_B T/mc^2\pi} \sim 1/20$ at $k_B T = 1 MeV$ and $\hat{v}/c \sim 1/6$ at $k_B T = 10 MeV$. The collective velocity$/c \leq$ incident velocity$/c \sim \sqrt{2E_{cm}/mc^2}$ where $E_{cm}$ is the center of mass energy per particle. However, the temperature is coupled to $E_{cm}$ via $3k_b T/2 \leq E_{cm}$. Thus $(u/\hat{v}) \sim 1$ or $\sim \frac{1}{2}$ the incident energy goes into flow and $\sim \frac{1}{2}$ into thermal energy. Under these conditions the Reynolds number is governed by $(d/l_\lambda)$. The concept of viscosity fails if $l_\lambda \gg d$. Taking $l_\lambda = d$, the Reynolds number is then $R_Y \sim (u/\hat{v}) \cdot (d/l_\lambda) \sim 1$. High Reynolds numbers occur when $d/l_\lambda \gg 1$. For $d/l_\lambda \sim 10$, $R_Y \sim 10$. Turbulence sets in when $R_Y \sim 10^3$. To get $R_Y \sim 10^3$ requires a very short mean free path or a very short relaxation time since $l_\lambda = <v> \tau_R$. But a short relaxation time destroys the flow, driving the system to thermal equilibrium in a collision. An analysis [1], based on a Fokker-Planck equation, of the time evolution of the momentum space density $f_P(\pm \vec{P}_0, t)$ of two colliding nuclei with initial momentum $\pm \vec{P}_0$ gave

$$f_P(\pm \vec{P}_0, t) = \frac{A \exp-(\left|\vec{P} \mp \vec{P}_0 e^{-\beta t}\right|^2 / 2mk_B T(1-e^{-2\beta t}))}{[2\pi m k_B T(1-e^{-2\beta t})]^{3/2}} \quad . \tag{53}$$

Each of the 2 colliding nuclei have $A$ nucleons in the overlapping fireball region. At $t=0$ $f_P(+\vec{P}_0, t) + f_P(-\vec{P}_0, t) = A(\delta(\vec{P}-\vec{P}_0) + \delta(\vec{P}+\vec{P}_0))$. The degradation of the centroid momentum is $\pm \vec{P}_0 e^{-\beta t}$, which is just the behavior of a particle started with $\pm \vec{P}_0$ and subject to a frictional force $-\beta \vec{P}$. The $1/\beta$ is a relaxation time. The variance of the momentum spreads with time as $k_B T(1-\exp(-2\beta t))$. An estimate of $t$ is the collision time of two overlapping nuclei which is the radius of the nuclei $R_A$ divided by the incident velocity $v_{inc}$ or $t_{coll} \sim R_A/v_{inc}$. Thus to have some persistence of the initial momentum, or collective motion, $\beta t_{coll} \sim 1 \sim \beta R/v_{inc}$. Using this last result and equating the relaxation time $1/\beta$ with the relaxation time $\tau_R$ that appears in the expression for the viscosity. The

$$R_Y \sim \frac{u}{<v>} \frac{v_{inc}}{<v>} \sim \frac{v_{inc}}{u} \frac{E_{coll}}{E_{th}} \quad . \tag{54}$$

where $E_{coll}$ is the collective energy and $E_{th}$ is the thermal energy of a particle.

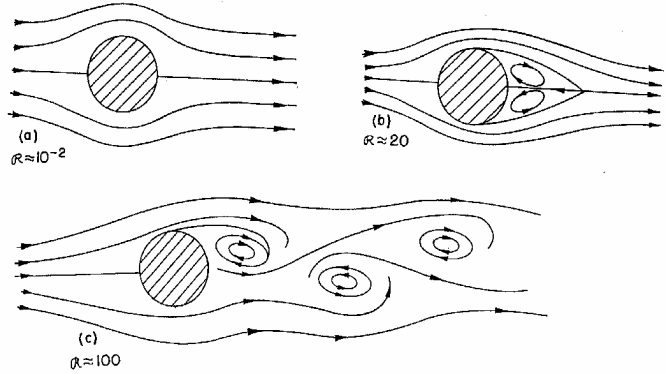

FIG. 8. Reynolds number $R_Y$ and flow. At low Reynolds number the flow past an obstacle is laminar. At higher $R_Y$ vortices appear behind the obstacle. At still higher $R_Y$ the vortices break off and flow with the fluid. Further discussion can be found in Feynman [52]

## V. Conclusions

Viscosity plays an important role in many areas of physics as discussed in the introduction. An initial study [1] showed that in a heavy ion collision viscosity reduced the collective flow [1]. Recent interest in the nature of hydrodynamic flow in RHIC experiments arose from the possibility that strongly coupled hadronic systems can behave as a perfect fluid [3-5]. Studies of nearly perfect fluids appear in atomic physics and in particular in ultracold Lithium atoms. Such studies employ the Feshbach resonance as a tuning device where an external magnetic field controls the scattering length and investigates the unitary limit where universal thermodynamics applies. Furthermore, a lower limit on the viscosity to entropy density ratio came from string theory [9] which has motivated considerable theoretical discussion and further experimental studies in several systems. Thus, a wide range of energy scales in very different systems are manifesting similar behavior. The present paper explored questions related the viscosity, the entropy and the ratio of viscosity to entropy density to see if a nuclear system at moderate energy or temperature behaves as a perfect fluid. The large experimentally observed scattering length in nucleon-nucleon systems makes this system very useful for studies of the unitary limit and universal thermodynamic behavior at moderate temperatures of several to tens of $MeV$ and at various densities.

The viscosity was studied in both a classical and quantum approach for several types of potentials. These include treating the collisions between nucleons as: A) billiard ball hard spheres scattering which is often used as a model in textbooks to discuss viscosity; B) interactions represented by an attractive square well tuned to the experimentally determined effective range and scattering length of nucleon-nucleon collisions; C) a combination of a short range repulsion and a longer range attraction which represents features associated with realistic interactions. The classical theory of the scattering angle was cast into a form that contains Snell's laws of reflection and refraction with an energy dependent index of refraction. The lowest classical value of the viscosity of the attractive potential is the hard sphere limit. The quantum theory involved calculation of the phase shifts from these potentials. A scaling law for the behavior of viscosity was shown to exist where the quantum calculation goes into the classical value in the limit that $\hbar \to 0$. This scaling law parallels a similar result in which Fraunhofer diffraction increases the hard sphere geometric cross section by a factor of two. The unitary limit for pure neutron matter, when correct for effective range factor, was found to hold over a broad range of temperatures. A simple analytic expression was also given for this system away from the unitary limit. Several other features that are unique to the nuclear system such as a tensor force and an inner hard core or strong repulsion are also mentioned. The inner hard core removes a divergence in $\eta$ as the index of refraction goes to unity.

The entropy and in particular the non-ideal gas interaction entropy was further developed beyond the initial results of Ref. [28,29]. Using results for the viscosity and entropy density, the ratio of the viscosity to entropy density was analyzed both in the unitary limit and away from the unitary limit. The importance of the isothermal compressibility in such studies was pointed out. The $\eta/s$ ratio was developed in two variables, the thermodynamic fugacity variable $z$ and a variable $y$ which involves the ratio of either the scattering length or interaction radius to interparticle distance. The minimum in the $\eta/s$ ratio in the unitary limit (approximately 10 times the string theory result of $(1/4\pi)\hbar/k_B$) was shown to occur at high fugacity and suggests that higher order correlations beyond the two particle case are necessary for a more accurate description. Calculation based on a hard sphere Fermi gas where shown to be several times larger than the unitary limit. The need to include higher order terms was also seen in the behavior of the isothermal compressibility which is negative at the minimum in $\eta/s$ for attractive interactions. Systems with negative values of the isothermal compressibility are mechanically unstable. The relation of these aspects with a van der Waal gas was developed and the properties associated with a liquid-gas phase transition and a critical point were noted. The results from RHIC suggest a much lower $\eta/s$ ratio and a more perfect liquid than the moderate energy nucleonic case considered here with a much higher $\eta/s$ ratio even in the unitary limit.

Finally, the importance of the viscosity in reducing the flow was illustrated using both a linear transport theory based on a relaxation time approximation to the Boltzmann equation and also using a Fokker Planck equation. The importance of the Reynolds number was stressed to see if the flow is laminar or turbulent.

**Appendix A.    Properties of a hard sphere gas**

Using $\tan \delta_l = j_l(x)/\eta_l(x)$, $x = kR_C$ and $d\delta_l/dx = -1/x^2(j_l^2(x)+\eta_l^2(x))$ the viscosity and entropy of a hard sphere gas can be calculated. Identical spin ½ fermions can coupled to total spin 0 in even orbital angular momentum states $l = 0,2,4,...$ and can couple to total spin 1 in odd $l = 1,3,5,...$ states. Spin 0 bosons only interact in even orbital angular momentum states. At low energies, the $S-$ wave interaction dominates and the $\omega$ integral is very accurately described by replacing the sum in $\phi$ with just the $l = 0$ term which is $2\sin^2(x)/3$. A dimensionless variable $y = \sqrt{2\pi}R_C/(V/A)^{1/3}$ is a measure of the interaction distance to particle separation. The $\omega$, including a factor of 2 for identical fermions or bosons and a factor $1/g_S^2 = 1/4$ for spin ½ fermions coupled to spin 0, is

$$\omega(\xi) = \frac{2}{g_S^2} 4\xi^4 \frac{2}{3} \frac{1}{8\xi^{11/2}} (\sqrt{\xi}(9\xi - 2) + \{4 + 5\xi(3\xi - 4)\} \cdot D(\frac{1}{\sqrt{\xi}})) \tag{A.1}$$

The $D(1/\sqrt{\xi})$ is the Dawson F function which has an expansion for large $\sqrt{\xi}$ given by the $D(\xi) = \xi^{-1/2} \Sigma_{n=0} (-1)^n 2^n / (\xi^n (2n-1)!!)$. An expansion in inverse powers of $\xi$ gives $\omega(\xi) \approx (2 \cdot 8/g_S^2) \cdot (1 - 4/3\xi + 8/9\xi^2 - 8/21\xi^3 + ...)$. The long wavelength limit for $S$-waves is obtained from $\sin^2 x \approx x$ giving $\omega(\xi) = 2 \cdot 8/g_S^2$. The $S_{int} = 0$ since $B_C(\xi) \sim 1/\xi^{1/2}$. The $\eta/s$ is then given by the equation

$$\frac{\eta}{s} = \frac{g_S^2}{2} \frac{\hbar}{k_B} \frac{5}{8} \sqrt{2\pi} \frac{2}{y^2} \frac{1}{z^{1/3}} \frac{1}{\omega(z^{2/3}/y^2)} \frac{1}{(5/2 - \ln z \pm z/(2^{7/2} g_S))}. \tag{A.2}$$

Table 1 summarizes properties of hard sphere Fermi and Bose gases in two extreme limits. Between these limits properties of these gases can be obtained from Eq. A.3-A.5.

---

TABLE 1. Limiting behaviors for the interaction entropy density, viscosity, ratio of viscosity to entropy density for spin ½ fermions $(g_S = 2)$ and spin 0 bosons $(g_S = 1)$.

---

| Variable | $S$ – wave long wavelength limit | Geometric short wavelength limit |
|---|---|---|
| $\dfrac{S_{int}}{V}$ | 0 | $-k_B \dfrac{A^2}{V^2} \dfrac{1}{2} \dfrac{4\pi R_C^3}{3}$ |
| $\eta$ | $g_S^2 \dfrac{5\sqrt{2\pi}}{128} \dfrac{\hbar}{\lambda_T} \dfrac{1}{\pi R_C^2}$ | $\dfrac{5\sqrt{2\pi}}{16} \dfrac{\hbar}{\lambda_T} \dfrac{1}{\pi R_C^2}$ |
| $\dfrac{\eta/s}{\hbar/k_B}$ | $\dfrac{g_S^2}{2} \dfrac{5\sqrt{2\pi}}{64} \dfrac{2}{y^2} \dfrac{1}{z^{1/3}(\frac{5}{2} + \ln g_S - \ln z \pm \frac{z}{2^{7/2} g_S})}$ | $g_S^2 \dfrac{5\sqrt{2\pi}}{16} \dfrac{2}{y^2} \dfrac{1}{z^{1/3}(\frac{5}{2} + \ln g_S - \ln z \pm \frac{z}{2^{7/2} g_S} - \frac{y^3}{3\sqrt{2\pi}})}$ |

---

The general result for $\eta/s$ as a function $(z, y)$ which includes all partial waves is

$$\frac{\eta}{s} = \frac{\hbar}{k_B} \frac{5\sqrt{2}\pi}{8} \frac{2}{y^2} \frac{1}{\omega(\xi)_{\xi=z^{2/3}/y^2}} \frac{1}{z^{1/3}\widehat{s}(\xi)_{\xi=z^{2/3}/y^2}} \tag{A.3}$$

where

$$\omega(\xi) = 2 \cdot 4\xi^4 \int_0^\infty \exp(-\xi x^2) x^7 \left( \frac{1}{x^2} \frac{1}{g_S^2} F_E(x) + \frac{(g_S-1)(g_S+1)}{g_S^2} F_O(x) \right) dx \tag{A.4}$$

$$F_E(x) = \sum_{l=0,2,\ldots} \frac{(l+1)(l+2)}{2l+3} \sin^2(\delta_{l+2} - \delta_l), \quad F_O(x) = \sum_{l=1,3,\ldots} \frac{(l+1)(l+2)}{2l+3} \sin^2(\delta_{l+2} - \delta_l)$$

and

$$\widehat{s}(\xi) = \frac{5}{2} + \ln g_S - \ln z \pm \frac{z}{2^{7/2} g_S} - 2^{3/2} \xi^2 \frac{d}{d\xi} \xi^{1/2} \widehat{B}_C(\xi) \tag{A.5}$$

$$\widehat{B}_C(\xi) = \frac{1}{\pi} \int_0^\infty \exp(-\xi x^2) \left( \frac{1}{g_S^2} M_E(x) + \frac{(g_S-1)(g_S+1)}{g_S^2} M_O(x) \right)$$

$$M_E(x) = \sum_{l=0,1,\ldots} (4l+1) \frac{d\delta_{2l}}{dx}, \quad M_O(x) = \sum_{l=0,1,\ldots} (4l+3) \frac{d\delta_{2l+1}}{dx}$$

The $P, D, F$ wave phase shifts for small $x$ are $\delta_1 = -x^3/3 + x^5/5 - x^7/7 + \ldots$, $\delta_2 = -x^5/45 + x^7/189 - \ldots$, $\delta_3 = -x^7/1575 + \ldots$. The $\delta_l = -x^{2l+1}/(2l+1)((2l-1)!!)^2$ is the leading order term for each $l$. The semi-classical limit has $\hbar \to 0$ and $\xi \to 0$. The $D-S$ contribution to $\omega(\xi)$ involves $\delta_0 - \delta_2 \sim x - x^5/45$, has $\sin^2(x - x^5/45) = x^2 - x^4/4 + 0x^6$ and is missing the lowest order $D-$wave interference term. Therefore, the $1/\xi^2$ term arises solely from the $P-$wave. The contribution of each partial wave is also given. It should be noted that the result of Eq.(18) gives an expansion for $\omega$ in inverse powers of $\hbar^2$ since $\xi = (\lambda_T/R_C\sqrt{2\pi})^2 \sim \hbar^2$. The viscosity is connected to this series expansion around the $S-$wave scattering limit using Eq.(4).

Some remarks regarding the entropy are as follows. A spinless Bose gas has only even $l-$terms. Therefore, the first contribution comes from a $D-$wave. Letting $S_{int,2}$ be the $D-$wave contribution to $S_{int}$ an expansion for large $\xi = \lambda_T^2/2\pi R_C^2$ gives

$$S_{int,2} = -k_B \frac{A^2}{V} \frac{20\pi^2 R_C^3}{3} \frac{R_C^2}{\lambda_T^2}. \tag{A.5}$$

A $P$-wave interaction for fermions gives an interaction entropy $S_{int,1}$ which is

$$S_{int,1} = -k_B \frac{A^2}{V} \frac{3}{g_S^2} 2^{3/2} (2\pi R_C^2)^{3/2} \frac{3}{\pi} \frac{\pi}{2} \xi^2 \left( \frac{e^\xi (1+2\xi)}{2\sqrt{\xi}} Erfc(\sqrt{\xi}) - \frac{1}{\sqrt{\pi}} \right) \tag{A.6}$$

Using $\xi = b/R_C^2 = \lambda_T^2 / 2\pi R_C^2$ and the small $x$ expansion of the phase shifts leads to an associated large $\xi$ low $T$ expansion of $B_C = B_{C,E} + B_{C,O}$. The $B_{C,E}, B_{C,O}$ are the even, odd $l$ parts of $B_C$. To order $1/\xi^{7/2}$:

$$\sqrt{\pi} B_C = -\left( \frac{1}{2\xi^{1/2}} + \frac{5}{24\xi^{5/2}} - \frac{25}{144\xi^{7/2}} \right)_E - \left( \frac{3}{4\xi^{3/2}} - \frac{9}{8\xi^{5/2}} + \frac{1}{\xi^{7/2}} (\frac{15}{16}(3+\frac{7}{225})) \right)_O \tag{A.7}$$

A comparison between the S-wave hard sphere Fermi gas and a Fermi gas based on all terms as developed in this appendix is shown in Fig. 9 for the case of $y = 0.4$

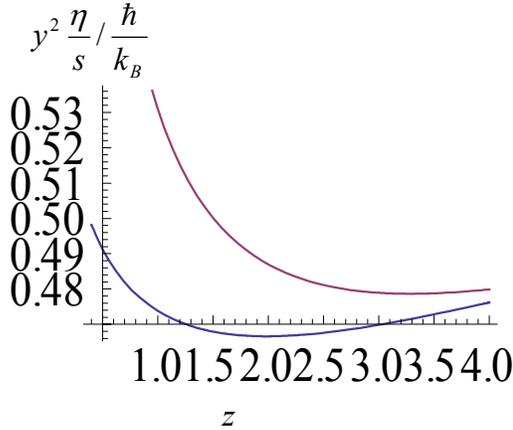

FIG. 9. (Color online) Comparison between the S-wave hard sphere Fermi gas and a Fermi gas based on all terms using Eq. (A.3). The results are for $y = 0.4$. The upper curve is the $S$-wave result. Some differences exist between the two curves, such as the shift to lower fugacity when higher partial waves are included. However the value at the minimum hasn't changed significantly. The value of $\eta/s = 3\hbar/k_B$ when $y^2\eta/s = .48\hbar/k_B$.

This work is supported by Department of Energy under Grant DE-FG02ER-409DOE and was done in part at Triumf Laboratory in Vancouver, BC.

___________________________________________